# AUGMENTED SEGMENTATION AND VISUALIZATION FOR PRESENTATION VIDEOS

*Alexander Haubold and John R. Kender*

Department of Computer Science, Columbia University, New York, NY 10027

{ahaubold,jrk}@cs.columbia.edu

## ABSTRACT

We investigate methods of segmenting, visualizing, and indexing presentation videos by separately considering audio and visual data. The audio track is segmented by speaker, and augmented with key phrases which are extracted using an Automatic Speech Recognizer (ASR). The video track is segmented by visual dissimilarities and augmented by representative key frames. An interactive user interface combines a visual representation of audio, video, text, and key frames, and allows the user to navigate a presentation video. We also explore clustering and labeling of speaker data and present preliminary results.

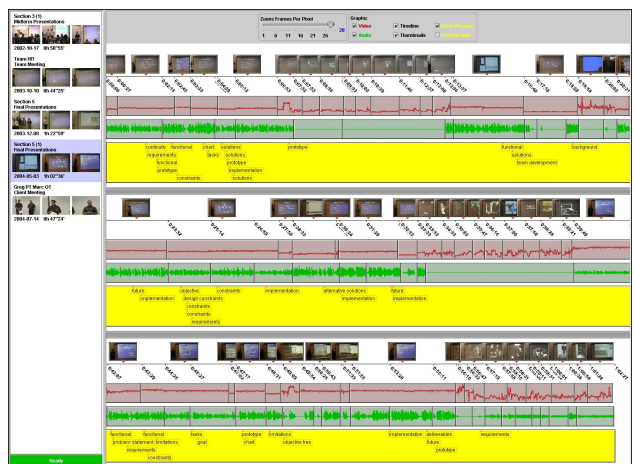

Figure 1: User Interface for video/audio segmentation and text augmentation. Videos are selected from the list on the left side. Video contents on the right side is shown on a segmented linear timeline, where one hour of video is compressed at 28 frames/pixel. The red graph (below timeline) shows video activity, the green graph (below red graph) audio activity, and the yellow graph (below green graph) displays index phrases.

## 1. INTRODUCTION

Video segmentation, visualization, and indexing have received much attention for providing means to organize and access video libraries. With the growing use of videos in classrooms other than for recording lectures, we investigate presentation videos. Characteristically, classroom presentations are carried out by several students and follow a known structure. Analysis of such videos should take advantage of this a priori knowledge.

Easily accessible presentation video libraries would allow for more efficient retrieval of specific presentation or speaker video clips. In a recorded classroom environment students would be able to quickly access and review their archived presentations and those from peers. Where videos were traditionally used as sequential streams with mere fast forward and reverse functions, they would now be laid out in a compressed summarizing interface for fast visual browsing.

Related work in video segmentation and summarization has focused mostly on lecture and news videos. In [1], lecture-style audio-video presentations are summarized in segments, which are 20-25% of the original video's length. Cues are taken from pauses in the audio track and slide transitions in the video track. Video skims [2] have been used to summarize news videos in much more compact representations taking cues from video, audio, and text. Approaches to structuring videos include determining topics and their temporal relevance in lecture videos [3], and finding story units in news video [4].

Our approach to segmenting and building an interactive user interface is based on separately considering audio and video data, and combining the resulting segmentations in a user interface. In addition, we generate an imperfect transcript using the IBM ViaVoice ASR, from which we filter a small number of meaningful phrases using text analysis presented in [5]. These phrases are used to index the video, and allow for quick visual scanning of a video's contents.

## 2. DEFINITION OF PRESENTATION VIDEOS

A presentation video contains one or more distinct presentations carried out by students or teams of students. Typically, an electronic medium like PowerPoint is used to accompany the speaker; however, for the purpose of segmentation and visualization, it is not required here. Contents are not limited to slides and speakers, and may include the screening of a short video clip, a discussion,

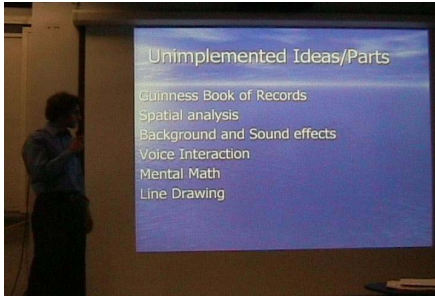 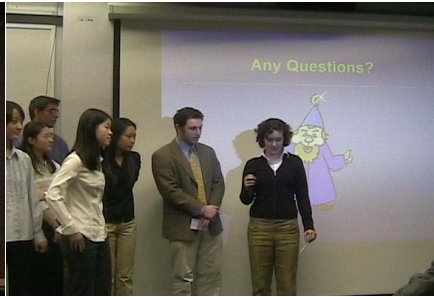 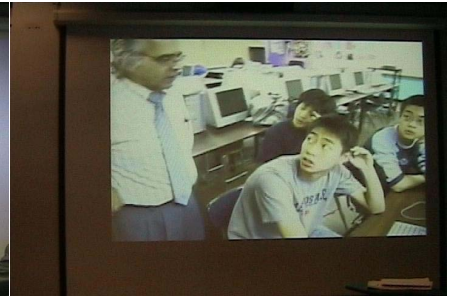

Figure 2a: Presentation       Figure 2b: Discussion and Q/A       Figure 2c: Film clip screening
Figure 2: Example camera shots from a presentation video.

etc. (see Figure 2).

A single camera captures the speaker standing next to the projected image, and a wireless microphone is used by the speaker to better pick up the audio signal. The time period during which a speaker presents does not necessarily overlap with the projected slides; two presenters may share a slide. Separating audio/video segmentation and visualization is especially useful for these conditions.

## 3. VIDEO SEGMENTATION

In this stage of segmentation, visual contents from a video is analyzed for shot boundaries. We apply methods of computing histogram changes between consecutive frames and detecting long-term changes by comparing the degree of change over time. Comparisons are made between two four-second windows, and a shot boundary is declared if the difference between the windows deviates significantly from the mean across both windows. An experimentally derived threshold is used to measure the deviation.

We have found this method to be robust in detecting changes in presentation slides. More interestingly, this method also detects speaker changes by differentiating the characteristic movement patterns between two speakers. We therefore found it important to include this measure in the user interface as a visual activity graph. (see Figure 3). It is also easy to visually pick out video segments with a high degree of visual change from the activity graph which in the example of Figure 3 (mid section) represents a film screening.

## 4. AUDIO SEGMENTATION

Audio segmentation for presentation videos lends itself to segmentation by speaker. We employ the method of detecting speaker changes via the Bayesian Information Criterion introduced in [6]. The audio track is sampled at regular intervals and vectors of 13 Mel Frequency Cepstral Coefficients are determined for each set of audio samples. Using a two-window approach, the BIC is computed for each partition of this interval.

If there exists a clear positive maximum among BIC values, a speaker change has been found, otherwise the interval is extended by the following audio samples.

In evaluating this method on presentation videos, we have found that the best segmentation is achieved with the following settings:

- Number of sample sets for which MFCC are computed in one second of audio $\approx 8$. This corresponds roughly to the number of syllables uttered in such a time frame.
- Length of each set of samples in terms of audio sample frequency $\approx \frac{f}{62.5}$. (i.e. 32kHz: 512 samples, 16kHz: 256 samples, 8 kHz: 128 samples). We thus extract a sound fragment approximately $\frac{1}{8}$ the length of a syllable and use it for the final BIC calculation.

Choosing the number of MFCC vectors per second much higher or lower, or selecting much longer, or shorter sample sets results in over- or under-segmentation.

The results from speaker segmentation are very favorable. From experiments we observed no false negatives, and only few false positives. The latter tend to be introduced by the occurrence of small pauses in the audio track. The final segmentation, as well as the audio activity graph (=audio amplitude) are included in the user interface.

## 5. TEXT AUGMENTATION

Parallel to visually summarizing video clips with thumbnails, we use text to summarize audio clips. However, transcripts are not readily available for the presentations, and we cannot make the assumption that every presentation is accompanied by electronic slides.

We thus generate transcripts using the IBM ViaVoice ASR. The resulting transcripts are highly imperfect with large Word Error Rates ($\approx 75\%$) due to several factors. Primarily, the audio quality varies greatly and depends on the individual presenter and the presentation environment. Due to the large number of speakers, it is impractical to apply speech model adaptation. Language model adaptation is also unfeasible, as the contents, style, and fluency of the presentations have high variance as well.

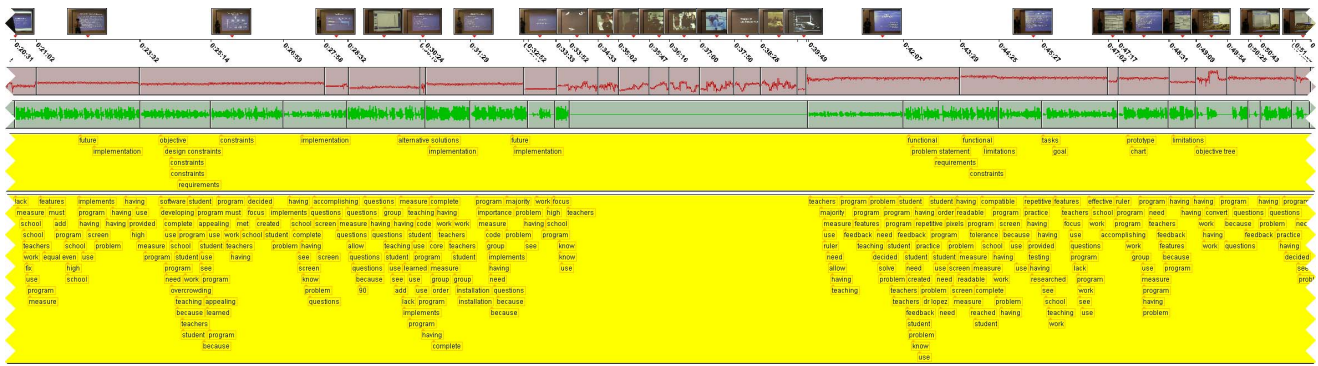

Figure 3: Complete timeline includes thumbnails for sufficiently long video segments (row 1), a timeline with time markers combining video and audio segmentations (row 2), visual video segmentation with activity graph (row 3: red), visual audio segmentation with activity graph (row 4: green), index phrases (row 5: yellow), text phrases (row 6: yellow).

| background | schedule | design constraints |
| limitations | tasks | alternative solutions |
| implementation | demo | functional requirements |

Table 1: Excerpt of topic phrases for presentations in the course "Engineering Design".

Transcripts with high error rates do not lend themselves to known text analyses, in which correlations are found between repetition and uniqueness of words and phrases. In a previous work [5], we have introduced methods by which highly imperfect transcripts from university lecture courses are filtered by using expected significant index terms extracted from external course-related sources such as textbooks, web pages, etc. We apply a similar method to transcripts from presentation videos. While we do not have indicators of the specific contents for a given presentation, we do have some knowledge about the overall structure. Presentations in the domain of our test video database revolve around Engineering Design projects. We have manually generated a list of 30 index phrases, which we expect to find in each presentation, and we use them to filter the transcript (see Table 1). The resulting "theme phrases" are included in the user interface and provide the equivalent of a table of contents for each presentation (see Figure 3, row 5).

Besides identifying theme phrases, we also apply text filtering of all of the phrases found in the source data of the electronic presentation slides, if available. To this end, each line of text in the slides is used as a phrase. The resulting "topic phrases" are included as an additional index in the user interface and give clues about specific items discussed in the presentation, including names, locations, numbers, etc.

## 6. INTERFACE

The interactive user interface is modeled as a linear time line (see Figure 3), sectioned to fit the screen (see Figure 1). This provides an overview of the presentation video's structure and contents.

Audio and video segmentations are included in the user interface as visually segmented activity graphs. The activity curve for audio represents audio amplitude, and for video the amount of change between two adjacent frames or clips. Activity on the video track is particularly interesting, as it provides clues about the amount of action at any given point. A more or less steady horizontal line indicates a video segment with conversational qualities, while a prolonged fluctuating line points out intense motion, e.g. in a film screening or an interactive demonstration.

A timeline combines separate audio and video segmentations. Thumbnails for sufficiently long video segments are placed above the timeline, and theme and topic phrases, if available, are placed below the audio segmentation graph.

For further exploration of the video, a zoom feature has been implemented that can be used to stretch the graph from 30 video frames/pixel to 1 frame/pixel. Clicking on thumbnails revels their original size. We plan on extending the interface to allow audio/video playback from any point in the graph.

While we have not conducted explicit user studies, the interface has been modeled after observations of instructors and students. The subjects in our classroom tend to have some familiarity with video editing, leading to the design of a row-media layout. We intend this view to be especially helpful for the extraction of film clips, while the text augmentation rows serve as search indices.

## 7. SPEAKER CLUSTERING AND LABELING

We have conducted preliminary experiments with speaker clustering and labeling. Previous work on broadcast news [7] shows successful results for classification of audio and clustering of speaker segments using trained Gaussian Mixture Models.

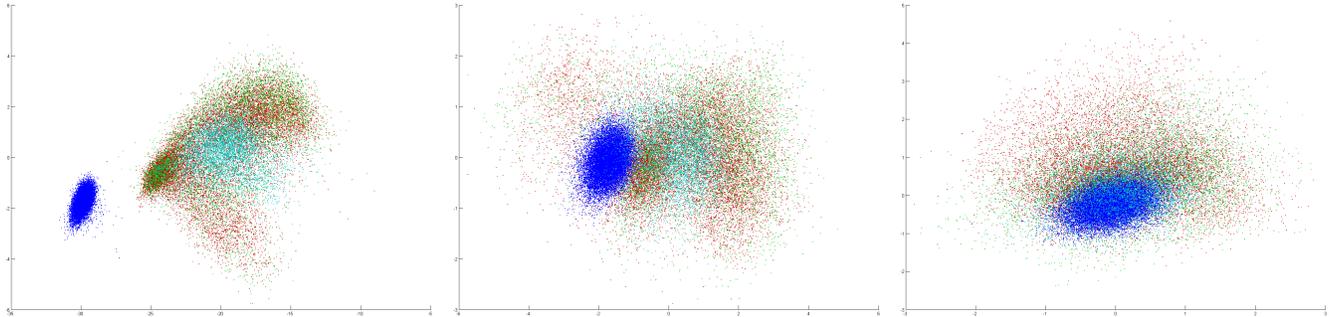

Figure 4a: MFCC 1 (x) vs. MFCC 2 (y)    Figure 4b: MFCC 2 (x) vs. MFCC 3 (y)    Figure 4c: MFCC 3 (x) vs. MFCC 4 (y)

Figure 4: MFC coefficients as discriminators of audio type measured for several short clips (45 minutes combined). ■ Female (4 speakers), ■ Male (6 speakers), ■ Film screening (1 clip), ■ Silence (4 segments). Except for the distinct cluster denoting silence, the three other types cannot be clearly distinguished among another. Lower cepstral coefficients have more discriminative power.

Using a bottom-up approach of combining speaker segments and evaluating the BIC for the new audio clip, we have identified several correct clustering matches, but also as many false positives. Some of these discrepancies are due to very short speaker segments (one to three seconds), and others are due to unclean audio tracks with ambient noise, such as the audience clapping.

Additional cues from video segmentation, namely the degree of motion by individual speakers, may help in increasing the accuracy of clustering. Audio segments for which the motion activity by the filmed speaker are similar have a higher likelihood of belonging to the same cluster. An example of speaker over-segmentation as a result of unclean audio data is included in Figure 3. The fragment underneath the two rightmost thumbnails displays two video and four audio segments. Since the video activity remains steady throughout three of the audio segments, it is very probable that the speaker has not changed.

In examining classification of audio segments with respect to female speech, male speech, film screening, and silence, we compared the discriminative power of the lower-order MFC coefficients for manually extracted audio clips (see Figure 4). We can clearly identify a distinct cluster for the audio type silence in Figure 4a. However, the types for female speech, male speech, and video screening are less distinctive. In fact, they overlap almost entirely. With increasing MFC coefficient, the distinctiveness between audio types becomes less clear. We are investigating the integration of speaker motion patterns with MFCC patterns to achieve a more reliable clustering.

## 8. CONCLUSION

We have presented methods of segmentation, text augmentation, and visualization of presentation videos. Our approach of separately analyzing and visualizing audio and video shows that the two media are neither inclusive nor exclusive, but complementary. We enhanced the segmentation by using index phrase filtering to provide further cues for visual browsing and searching of presentation video content.

In the immediate future, we plan to introduce our browsing tool to the classroom, for use and evaluation by students. We also intend to further investigate speaker clustering and labeling for the purpose of extracting additional structure from presentation videos.